\documentclass[traditabstract]{aa}
\usepackage{graphicx}
\usepackage{txfonts}
\usepackage{hyperref}
\hypersetup{
 colorlinks   = true, 
 urlcolor     = blue, 
 linkcolor    = blue, 
 citecolor   = blue 
 }

\newcommand{\m}{$^{\rm m}\!\!.$}

\newcommand{\ms}{M$_{\odot}$}
\newcommand{\rs}{R$_{\odot}$}
\newcommand{\mj}{M$_{\rm Jup}$}

\newcommand{\oc}{$O-C$}

\newcommand{\ond}{Ond\v{r}ejov}

\newcommand{\valmez}{Vala\v{s}sk\'e Mezi\v{r}\'{\i}\v{c}\'{\i}}

\newcommand{\vev}{Veversk\'a B\'it\'y\v{s}ka}

\begin{document}

\title{Possible substellar companions in dwarf eclipsing binaries:
\thanks{Based on new photometric observations from 
\ond\ observatory in the Czech Republic. 
Tables \ref{m1435}, ~\ref{m782}, and~\ref{m1425} are available 
in electronic form at {\tt www.aanda.org}.}}

\subtitle{SDSS~J143547.87+373338.5,
          NSVS~7826147, and
          NSVS~14256825
          }

\author{ M. Wolf~\inst{1}
        \and H. Ku\v{c}\'akov\'a~\inst{1,2,3}
        \and P. Zasche~\inst{1}
        \and K. Hornoch~\inst{2} 
        \and J. K\'ara~\inst{1}
        \and J. Merc~\inst{1,4}
        \and M.~Zejda~\inst{5}
                              }
\institute{Astronomical Institute, Faculty of Mathematics and Physics, Charles University, V~Hole\v{s}ovi\v{c}k\'ach~2, CZ-180~00~Praha~8, \\
           Czech Republic, \email{wolf@cesnet.cz}
\and Astronomical Institute, Academy of Sciences, Fri\v{c}ova~298,
       CZ-251~65~Ond\v{r}ejov, Czech Republic  
\and Research Centre for Theoretical Physics and Astrophysics, 
Institute of Physics, Silesian University in Opava, 
Bezru\v{c}ovo n\'am. 13, CZ-746~01 Opava, Czech Republic 
\and Institute of Physics, Faculty of Science, P. J. \v{S}af{\'a}rik University, Park Angelinum 9, 040 01 Ko\v{s}ice, Slovakia
\and Department of Theoretical Physics and Astrophysics, Masaryk University, Kotl\'a\v{r}sk\'a 2, CZ-611~37 Brno, Czech Republic
 }
             
\date{Received \today}

\abstract{We present the new results of our long-term observational project to detect the small variations in the orbital periods of low-mass and short-period eclipsing binaries. 
About 120 new precise mid-eclipse times 
were obtained for three relatively well-known dwarf eclipsing binaries: 
            SDSS~J143547.87+373338.5~($P=0.126$~d),
            NSVS~07826147~(0.162~d), and 
            NSVS~14256825~(0.110~d).
Observed-minus-calculated (O-C) diagrams of these systems were analyzed using all accurate timings, and, where possible, new parameters of the light-time effect were calculated.
For the first time, we derive (or improve upon previous findings with regard to) the short orbital periods of 13 and 10 years of possible third bodies for SDSS~J143547.87+373338.5 and NSVS~07826147, respectively. In these binaries, our data show that period variations can be modeled simply on the basis of a single circumbinary object.
For the first two objects, we calculated the minimum mass of the third components to be 17 \mj\, and 1.4 \mj\ , respectively, which corresponds to the mass of a brown dwarf or a giant planet. 
For NSVS~14256825, the cyclical period changes caused by a single additional body cannot be confirmed by our recent eclipse time measurements. 
More complex behavior connected with two orbiting bodies, or yet unknown effects, should be taken into account. 
 }


\maketitle

\section{Introduction}

Eclipsing binaries (EB) with a white dwarf or hot subdwarf component belong to rather curious stellar systems. Their typical light curves with a deep and narrow primary minimum and strong reflection effect are clear sign allowing the simple identification of these unique objects. The difference between primary and secondary surface temperatures is typically 25~000 -- 30~000~K. 
Short orbital periods of up to 0.5 days are very sensitive to any changes caused possibly by a mass transfer between components, magnetic field changes in a late-type secondary, or the presence of an unseen third body. Moreover, many low-mass stars
often show phenomena associated with magnetic activity, such as flares and star spots.

The small size of the binary components enables us to determine the eclipse times of this type of binary system with high precision (up to seconds). Therefore, very small amplitude variations in the orbital period can be detected by analyzing the observed-minus-calculated (O-C) diagram or eclipse-time-variation (ETV) curve. This makes them very promising targets in the search for circumbinary brown dwarfs or giant planets by analyzing the light-time effect (LITE). 
Several discoveries of planetary-mass companions orbiting the post-common envelope binaries (PCEB) and cataclysmic variables (CV) were announced in the past; for example,
RR~Cae \citep{2012MNRAS.422L..24Q},
DE CVn \citep{2018ApJ...868...53H}, 
HW~Vir \citep{2012A&A...543A.138B}, and
HS~0705+6700 \citep{2012A&A...540A...8B, 2013MNRAS.436.1408Q}.

The origin of dwarf binaries and their multiple systems 
is still an unresolved question in star formation theory. 
The discovery of circumbinary objects, planets, or brown dwarfs, as well as their identification and characterization is also highly relevant to recent exoplanet studies in low-mass multiples \citep{2013MNRAS.429L..45P, 2014MNRAS.438..307H}.

Here, we report on a long-term mid-eclipse time campaign of three similar EBs containing a subdwarf or white dwarf primary component (sdB or WD).
All these systems are relatively well-known northern hemisphere objects whose uninterrupted observations last almost 20 years.
Their short orbital periods are up to four hours, and important spectroscopic observations have been published for all of them.
This paper is a continuation of our previous period study of low-mass eclipsing binaries presented in \cite{2016A&A...587A..82W, 2018A&A...620A..72W}.

\begin{table*}
\caption{Observational log of selected eclipsing binaries.} 
\label{obs}  
\begin{tabular}{llccccccc}
\hline\hline\noalign{\smallskip}
System & Abbreviation & Type & Observed & Exposure & Filter & Number & Number \\
    & used in paper &   &  since   & time [s] &       & of frames & of minima  \\
\noalign{\smallskip}\hline
\noalign{\smallskip}          
SDSS J143547.87+373338.5 & S1435 & PCEB  & Apr 2012 & 30 & C & 1210 &  30 \\
NSVS 07826147            & N782  & sdB+M & Feb 2012 & 30 & R & 2810 &  50 \\
NSVS~14256825            & N1425 & PCEB  & Jul 2009 & 30 & R & 2118 &  41 \\
\noalign{\smallskip}\hline
\end{tabular}
\end{table*}

\section{Photometry of eclipses}

Since 2009, we have accumulated over 6~000 photometric
measurements mostly during primary eclipses and derived  
121 new precise times of minimum light for all three systems. 
The CCD photometry was obtained primarily at the \ond\ Observatory, Czech Republic, using the Mayer 0.65-m ($f/3.6$) reflecting telescope with the CCD camera G2-3200 and photometric R filter, or without a filter.
Such a long-term monitoring campaign with identical equipment is not frequent in current photometrical surveys (see Table~\ref{obs} for details of our observations).
A standard calibration (dark frame, flat field) was applied to the obtained CCD frames. The {\sc Aphot}, a synthetic aperture photometry and astrometry software, was routinely used for our time series. Alternatively,
{\sc C-Munipack}\footnote{\url{http://c-munipack.sourceforge.net/}.}
was used by observers to reduce time series over several nights.
Differential photometry was carried out using selected nearby comparison and check stars.
Concerning the other photometric procedures, observational circumstances, and data handling (e.g., time synchronization during observation, accurate mid-eclipse time determination, conversion to the barycentric Julian date dynamical time (BJD$_{\rm TDB}$), and adopted weighting of individual times), we invite the reader to consult our last paper, \cite{2018A&A...620A..72W}.

\section{Eclipse time variations} 

An orbiting circumbinary body in an eclipsing binary can be detected by the well-known light-time effect (LITE) as a result of delays or advances in the timings of a minimum light.
This effective tool was historically introduced by \cite{1952ApJ...116..211I, 1959AJ.....64..149I}, who also described a simple fitting procedure for the elements of the light-time orbit. 
To calculate the LITE, the suitable equations were presented by \cite{1990BAICz..41..231M}.
An interesting review on many applications of popular O-C diagrams in various astrophysical contexts can be found in \citet{2005ASPC..335.....S}.

There are seven independent variables 
to be determined in this procedure. These are as follows: 
the orbital period of the binary, $P_s$, 
the orbital period of the third body, $P_3$,
the semi-amplitude of LITE, $A$, 
the eccentricity of the outer orbit, $e_3$, and
the periastron passage time of the third body, $T_3$. 
The zero epoch is given by $T_0$, and the corresponding position 
of the periastron of the third-body orbit is given by $\omega_3$.

\begin{table}
\caption{New minima timings of S1435.} 
\label{m1435}
\begin{tabular}{clcc}
\hline\hline\noalign{\smallskip}
BJD$_{\rm TDB}$ -- & Error & Epoch  & Weight   \\
24 00000          & [day] &        &          \\
\noalign{\smallskip}\hline
\noalign{\smallskip}
   56026.386088 &  0.0001  & 14946.0 & 5   \\   
   56052.517278 &  0.0001  & 15154.0 & 5   \\
   56055.406808 &  0.0001  & 15177.0 & 5   \\ 
   56141.338248 &  0.0001  & 15861.0 & 5   \\
   56398.379163 &  0.00001 & 17907.0 & 10  \\   
   56436.445252 &  0.0001  & 18210.0 & 5   \\   
   56642.731207 &  0.00001 & 19852.0 & 10  \\
   56711.451305 &  0.00001 & 20399.0 & 10  \\
   56718.486595 &  0.00001 & 20455.0 & 10  \\
   56827.408563 &  0.00001 & 21322.0 & 10  \\   
   57467.498042 &  0.0001  & 26417.0 & 5   \\
   57482.573733 &  0.00001 & 26537.0 & 10  \\
   57516.368333 &  0.0001  & 26806.0 & 5   \\
   57531.444162 &  0.00001 & 26926.0 & 10  \\
   57725.669531 &  0.00001 & 28472.0 & 10  \\
   57780.570243 &  0.00001 & 28909.0 & 10  \\
   57980.323354 &  0.00001 & 30499.0 & 10  \\
   58171.533757 &  0.00001 & 32021.0 & 10  \\
   58270.405389 &  0.0001  & 32808.0 & 5   \\
   58337.366630 &  0.00001 & 33341.0 & 10  \\ 
   58529.456525 &  0.00001 & 37870.0 & 10  \\
   58530.712795 &  0.0001  & 34880.0 & 10  \\
   58532.471465 &  0.00001 & 34894.0 & 10  \\ 
   58550.688076 &  0.0001  & 35039.0 & 5   \\
   58627.574378 &  0.00001 & 35651.0 & 10  \\
   58663.504759 &  0.00001 & 35937.0 & 10  \\
   58933.360318 &  0.0001  & 38085.0 & 5   \\
   58976.451720 &  0.0001  & 38428.0 & 5   \\
   59074.318353 &  0.00002 & 39207.0 & 10  \\
   59215.527737 &  0.00002 & 40331.0 & 10  \\
\noalign{\smallskip}\hline
\end{tabular}
\end{table}

Generally, the LITE method is more sensitive to companions on a long-period orbit, and its semi-amplitude is proportional to the mass and period of the third body as
$$ A \sim M_3\ P_3^{2/3}. $$
Moreover, low-mass binary components favor the detection of low-mass companions on short-period orbits \citep{2012AN....333..754P}.
On the other hand, in the case of a shorter orbital period of the third body 
(usually less than one year), small dynamical perturbations of the inner orbit can occur that also create additional changes in the observed times
(see \cite{2011A&A...528A..53B, 2016MNRAS.455.4136B}).

\subsection{SDSS J143547.87+373338.5}

The detached eclipsing binary SDSS~J143547.87+373338.5
(also WD~1433+377, $G$ = 17\m0, Sp. DA+M4.5, GAIA parallax $5.45\pm 0.07$ mas) is rather faint but relatively well-studied object with a short orbital period of about three hours. It belongs to a post-common envelope binary (PCEB) type, which contains a white dwarf primary and a red dwarf secondary. As mentioned in the literature, S1435 is probably a pre-CV system just at the upper edge of the known period gap of cataclysmic variables.
Short eight-minute eclipses were discovered by \cite{2008ApJ...677L.113S}, who also estimated the first intervals of preliminary absolute parameters: 
$M_1$ = 0.35–0.58~\ms, $R_1$ = 0.0132–0.0178~\rs, 
$M_2$ = 0.15–0.35~\ms\ and $R_2$ = 0.17–0.32~\rs. 

The spectroscopic parameters were later improved by 
\cite{2009MNRAS.394..978P}, who found a consistent set of absolute elements:
$M_1$ = 0.48–0.53~\ms, $R_1$ = 0.014–0.015~\rs, 
$M_2$ = 0.19–0.25~\ms\ and $R_2$ = 0.22–0.25~\rs.
The last period study of S1435 was presented by \cite{2016ApJ...817..151Q}, who
announced a rapid decreasing of the orbital period at a rate of about $-8 \cdot 10^{-11}$ s s$^{-1}$. As an alternative scenario, they also proposed the LITE caused by an unseen brown dwarf orbiting the eclipsing pair with the period of 7.72~years.

Our observations presented here cover the time span of about 25~000 epochs, which corresponds to 8.5~years. 
Using our newly derived eclipse times listed in Table~\ref{m1435} together with those obtained by \cite{2008ApJ...677L.113S}, \cite{2009MNRAS.394..978P}, and \cite{2016ApJ...817..151Q}, we improved the LITE elements given in Table~\ref{t2}. The following linear light elements were used for epoch calculation:

\begin{center}
Pri.Min. = BJD 24 54148.70395(2) + 0.125630981(5) $\cdot\ E$.
\end{center}

\begin{figure}[t]
\centering
\includegraphics[width=\columnwidth]{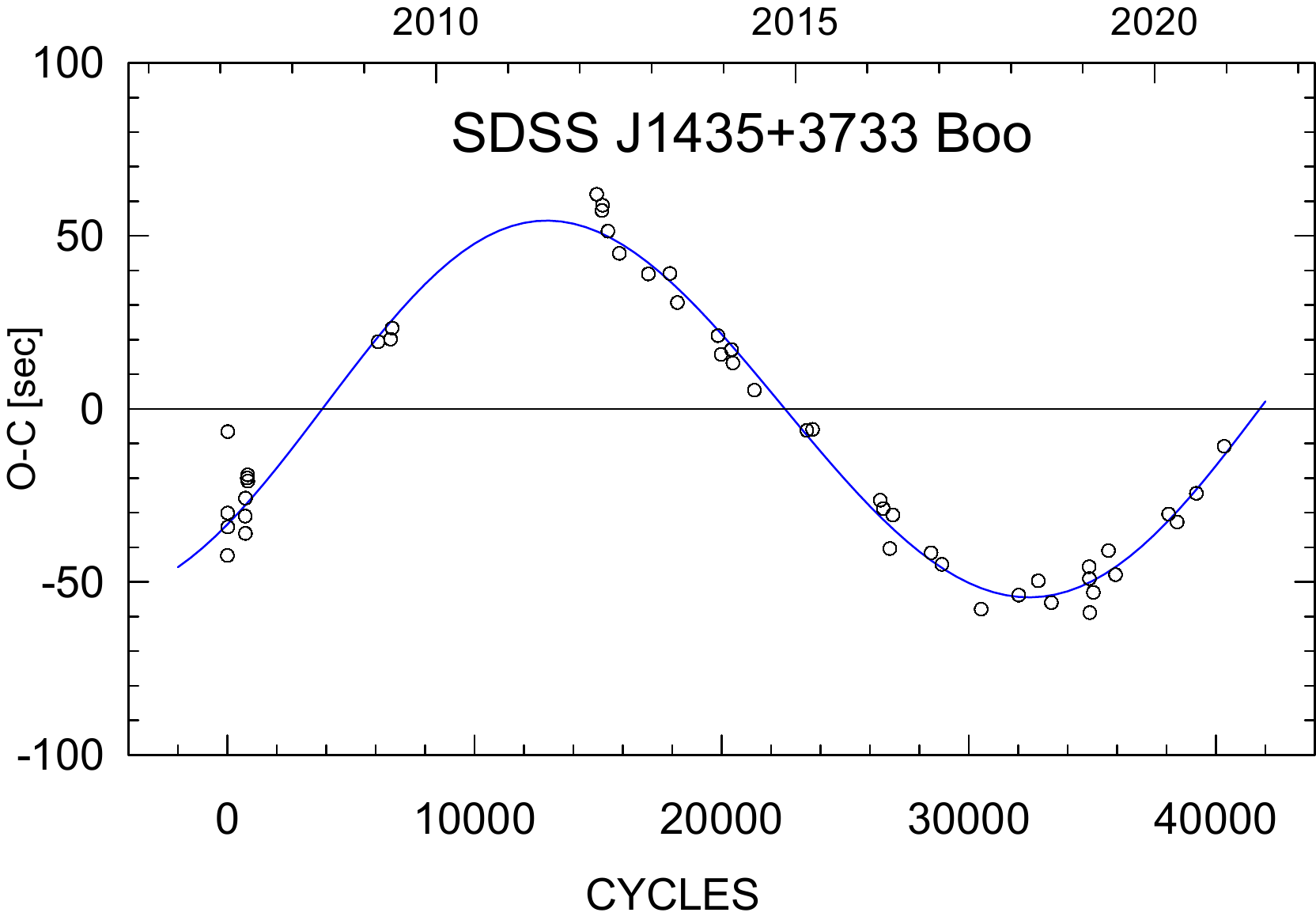}
\caption[]{Current O-C diagram for the eclipse times of S1435. 
        The blue sinusoidal curve represents the LITE with the short period of about 13~years and a well-defined semi-amplitude of 54~sec.
        The individual primary minima are denoted by circles.
     }
\label{s1435}
\end{figure}

\noindent
A total of 48 accurate mid-eclipse times of primary minimum were included in our analysis. The computed LITE parameters and their internal errors of  the least-squares fit are given in Table~\ref{t2}.
The current O-C diagram is plotted in Fig.~\ref{s1435}, where the sinusoidal trend is clearly visible. The nonlinear prediction, corresponding to the fit parameters, is plotted as a continuous blue curve.
One whole orbital period of the possible third body is now covered by CCD measurements.

\begin{table*}
\caption{LITE parameters for S1435 and N782  
         (with errors of the last digit in parentheses). }
\begin{tabular}{cccccc}
\hline\hline\noalign{\smallskip}
Element & Unit      &    S1435       & N782        \\ 
\noalign{\smallskip}\hline\noalign{\smallskip}
$T_0$ & BJD-2400000 & 54148.70395(2) & 55611.92657(3) \\
$P_s$ & days        & 0.125630981(5) & 0.161770446(2) \\
$P_3$ & days        & 4765(85)       & 3820(140)     \\
$P_3$ & years       & 13.0(0.3)      & 10.5(0.4)     \\ 
$e_3$ &  --         & 0.05(4)       & 0.0           \\ 
$A$   & days        & 0.00063(2)     & 0.000050(3)   \\ 
$A$   & sec         & 54.4(1.7)      & 4.3(0.3)      \\ 
$\omega_3$ & deg    & 23.7(3.0)      & 204.1(2.5)   \\ 
$T_3$ &  JD-2400000 & 54930(20)      & 50155(20)    \\ 
$a_{12}\sin i$ & au &  0.109         & 0.0087       \\ 
$\sum{w\ (O-C)^2}$ & day$^2$ & $ 3.4\cdot10^{-7}$ & $1.4\cdot10^{-7}$ \\ 
\noalign{\smallskip}\hline
\end{tabular}
\label{t2}
\end{table*}

\subsection{NSVS 07826147 CrB = DD CrB }

\begin{figure}[t]
\centering
\includegraphics[width=\columnwidth]{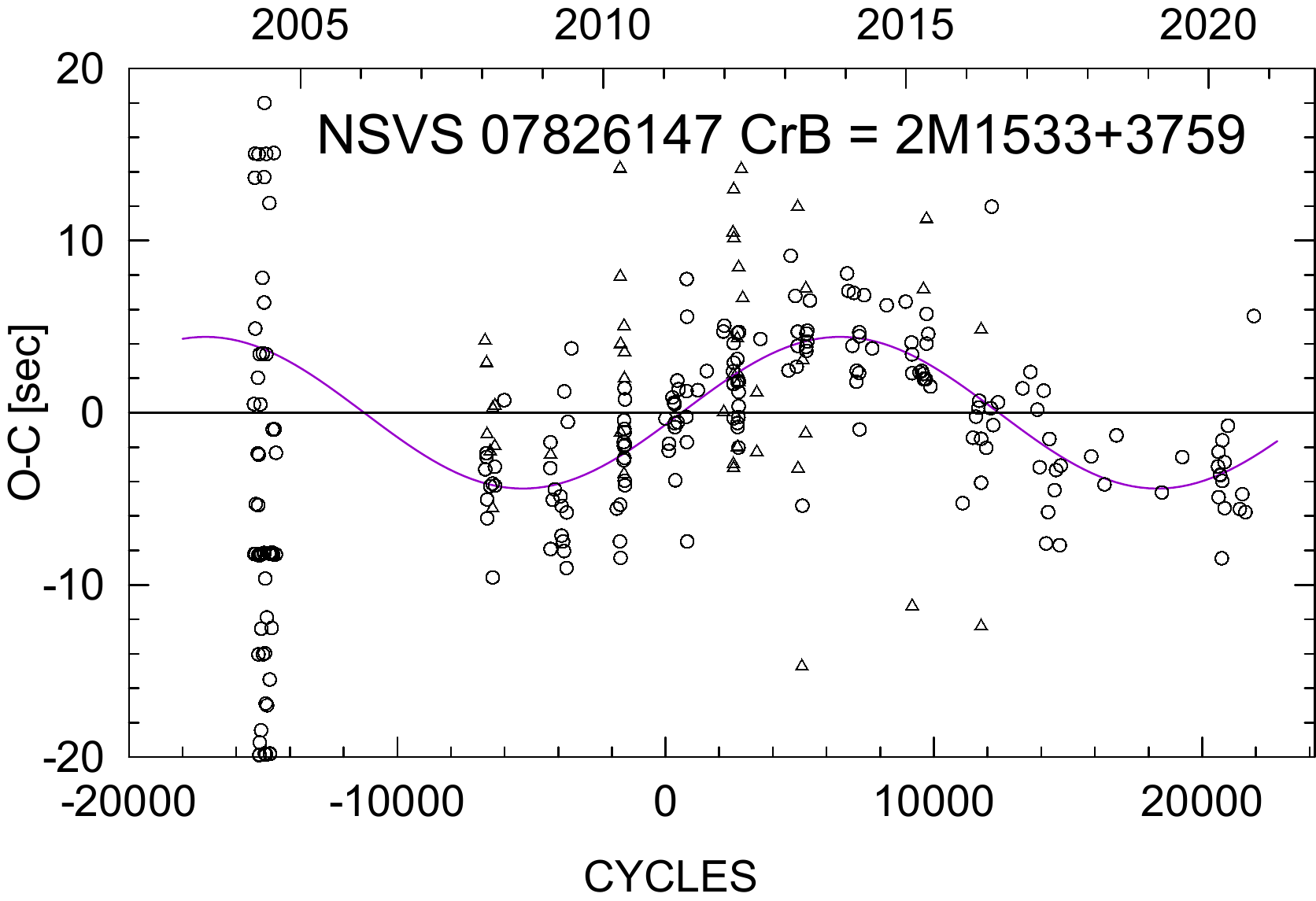}
\caption[]{Present O-C diagram for the eclipse times of N782. The sinusoidal curve represents the LITE with the short        period of about ten years and a very small semi-amplitude 
                of 4.3 sec. The individual primary minima are denoted by circles, the secondary by triangles. The first group of mid-eclipse times close to the cycle $-15~000$ derived by \cite{2014A&A...566A.128L} from the SuperWASP database was not taken in consideration 
                due to the large scatter of these data.}
\label{n782}
\end{figure}

\begin{table}[h!]
\caption{New minima timings of N782. } 
\label{m782}
\begin{tabular}{llrc}
\hline\hline\noalign{\smallskip}
BJD$_{\rm TDB}$ -- & Error & Epoch  & Weight   \\
24 00000          & [day] &        &          \\
\noalign{\smallskip}\hline
\noalign{\smallskip}
  55960.703707 &  0.00001 & 2156.0 & 10 \\
  55963.534637 &  0.0002  & 2173.5 & 0  \\
  55966.527447 &  0.00001 & 2192.0 & 10 \\
  56055.339417 &  0.00001 & 2741.0 & 10 \\
  56077.421107 &  0.0002  & 2877.5 &  0 \\  
  56162.350527 &  0.0002  & 3402.5 &  0 \\  
  56162.350487 &  0.0002  & 3402.5 &  0 \\  
  56184.270457 &  0.00001 & 3538.0 & 10 \\  
  56366.424034 &  0.00001 & 4664.0 &  5 \\  
  56394.410294 &  0.00001 & 4837.0 & 10 \\  
  56461.383233 &  0.00001 & 5251.0 & 10 \\
  56482.413413 &  0.00001 & 5381.0 & 10 \\  
  56706.627268 &  0.00001 & 6767.0 & 10 \\    
  56714.554008 &  0.00001 & 6816.0 & 10 \\  
  56738.495997 &  0.00001 & 6964.0 & 10 \\   
  56747.393407 &  0.00001 & 7019.0 & 10 \\   
  56809.351486 &  0.00001 & 7402.0 & 10 \\  
  56858.367895 &  0.00001 & 7705.0 & 10 \\  
  56945.238653 &  0.00001 & 8242.0 & 10 \\  
  57059.610360 &  0.00001 & 8949.0 & 10 \\   
  57185.467748 * & 0.00001 & 9727.0 & 10 \\
  57503.346624 &  0.00001 & 11692.0 & 10 \\  
  57764.605901 &  0.00001 & 13307.0 & 10 \\  
  57853.417861 &  0.00001 & 13856.0 & 10 \\  
  57892.404551 * & 0.00001 & 14097.0 & 10 \\
  57918.449511 &  0.00001 & 14258.0 & 10 \\  
  57926.376312 &  0.00001 & 14307.0 & 10 \\  
  57957.436203 &  0.00001 & 14499.0 & 10 \\  
  57966.333591 &  0.00001 & 14554.0 & 10 \\  
  57988.334321 &  0.00001 & 14690.0 & 10 \\  
  57994.319881 &  0.00001 & 14727.0 & 10 \\  
  58178.576424 &  0.00001 & 15866.0 & 10 \\  
  58258.491005 &  0.00001 & 16360.0 & 10 \\  
  58330.478886 &  0.00001 & 16805.0 & 10 \\  
  58604.356211 &  0.00001 & 18498.0 & 10 \\   
  58727.301773 &  0.00001 & 19258.0 & 10 \\   
  58942.618229 &  0.00001 & 20589.0 & 10 \\  
  58946.338959 &  0.00001 & 20612.0 & 10 \\  
  58946.500699 &  0.00001 & 20613.0 & 10 \\  
  58955.88340 \dag & 0.00001 & 20671.0 & 10 \\ 
  58956.04517 \dag & 0.00001 & 20672.0 & 10 \\ 
  58965.4278 **    &  0.0001 & 20730.0 & 5  \\   
  58968.33972 \dag & 0.00001 & 20748.0 & 10 \\ 
  58969.31037 \dag & 0.00001 & 20754.0 & 10 \\ 
  58981.92845 \dag & 0.00001 & 20832.0 & 10 \\ 
  58982.25196 \dag & 0.00001 & 20834.0 & 10 \\ 
  59074.461113 &  0.00001 & 21404.0 & 10 \\  
  59089.344004 &  0.00001 & 21496.0 & 10 \\  
  59108.271134 &  0.00001 & 21613.0 & 10 \\ 
  59159.228956 \ddag & 0.00001 &  21928.0 & 10 \\ 
\noalign{\smallskip}\hline
\end{tabular}
\tablefoot{ * \valmez, ** \vev, and \ddag\ MUO \\
observatories, Czech Republic, \dag\ \textit{TESS} photometry.}
\end{table}

The detached eclipsing binary NSVS~7826147 (also DD~CrB, FBS~1531 +381, 2MASS~J15334944 +3759282, CSS~6833, 
$V_{\rm max}$ = 13\m08, Sp. sdB+dM, GAIA parallax $1.9\pm 0.05$ mas) 
is a well-known northern and low-mass binary system 
with a very short orbital period ($P=0.16$ d). 
It was mentioned originally in the First Byurakan Spectral Sky Survey
\citep{1990Afz....33..213A} as a blue stellar object.
The eclipsing nature of the system was discovered by 
\cite{2007JSARA...1...13K} in the Northern
Sky Variability Survey \citep[NSVS;][]{2004AJ....127.2436W}.
\cite{2010Ap&SS.329..107L} measured 16 additional times of minimum and improved the orbital period of this binary ($P$ = 0\fd16177046).
The first photometric and spectroscopic analysis of N782 was presented by \cite{2010ApJ...708..253F}, who derived the precise physical parameters of both components: the sdB primary mass is 
$M_1= 0.376\pm0.055$ \ms\ and its radius is $R_1= 0.166\pm0.007$ \rs, and 
the secondary has $M_2= 0.113\pm0.017$ \ms, $R_2= 0.152\pm0.005$ \rs, consistent with a main-sequence M5~star.
\cite{2012A&A...538A..84B} determined a further seven times of minimum, and later \cite{2014A&A...566A.128L} provided additional timings from the SuperWASP database, extending the time span back to 2004.

\cite{2015PKAS...30..289Z} were probably first to announce the detection of a cyclical change in the period of this system; this was confirmed by 
\cite{2015AcPPP...2..183Z}, who reported that this periodic change could be caused by an unseen circumbinary object of mass greater than 4.7 \mj\ with an orbital radius of 0.64~au, introducing a LITE effect of 0.00004 days (3.5 s). Neither publication states a period, but
\cite{2015AcPPP...2..183Z} suggested 11~000 cycles, equivalent to 4.9~years. 
The last study of N782 based on long term photometry and next timings was presented by \cite{2017ApJ...839...39L}. They
also derived the precise absolute parameters of both components:
$M_1$ = 0.442(12)~\ms, $R_1$ = 0.172(2)~\rs, 
$M_2$ = 0.124(5)~\ms\ and $R_2$ = 0.157(2)~\rs.
They concluded that the orbital period of the system had remained constant for the past 12~years.
Finally, this interesting PCEB object was also included in the {\sc MUCHFUSS} photometric campaign \citep{2018A&A...614A..77S} and a new period study of \cite{2018A&A...611A..48P},
who also concluded that no significant variations are visible on the O-C diagram. The following linear light elements were derived in the last mentioned paper:

\begin{center}
Pri.Min. = BJD 24 55611.92655(1) + 0\fd161770449(2) $\cdot\ E$.
\end{center}

\noindent
All our new eclipse times are listed in Table~\ref{m782}.
In addition, we used the high-precision data obtained by the Transiting Exoplanet Survey Satellite 
\citep[TESS,][]{2015JATIS...1a4003R}. Our target was observed 
in Sector~24 in 2-minute cadence mode during April and May 2020. We derived six new times from the beginning, middle, and the end of this period. They are also listed in Table~\ref{m782}. All TESS minima perfectly fit the \oc\ curve.
Four additional mid-eclipse times of N782 were observed at 
\valmez, \vev\ and Masaryk University observatories in the Czech Republic.

All together, 284 reliable times of primary minimum light were included 
to our analysis, and the shallow and less precise secondary eclipses were not included due to a large scatter. The O-C diagram is shown in Fig.~\ref{n782},
and the computed LITE parameters and their internal errors of 
the least-squares fit are given in Table~\ref{t2}.
The nonlinear prediction, corresponding to the fit parameters, 
is plotted as a continuous violet curve in Fig.~\ref{n782}.

\subsection{NSVS 14256825 = V1828 Aql}

 The third eclipsing binary NSVS~14256825 (also V1828~Aql, 
 2MASS~J20200045+0437564, UCAC2~33483055, USNO-B1.0 0946-0525128, 
 $G$ = 13\m2, Sp. sdOB+M, GAIA parallax $1.1 9\pm 0.06$ mas) 
 is one of the well-known HW Vir-type systems with a short orbital period of 2.65 hours containing a very hot subdwarf B or OB primary. Its light variability was found in the NSVS data \citep{2004AJ....127.2436W}.  
 \cite{2007IBVS.5800....1W} performed the multicolor CCD observations and derived the first mid-eclipse times of N1425. Later, \cite{2012MNRAS.423..478A} analyzed multicolor photometry and the radial velocity curve simultaneously using the Wilson-Devinney code, and they provided the following fundamental parameters of N1425:
$M_1 = 0.419 \pm 0.070$ \ms, $M_2 = 0.109 \pm 0.023$ \ms, 
$R_1 = 0.188 \pm 0.010$ \rs, $R_2 = 0.162 \pm 0.008$ \rs, 
and $i = 82.5 \pm 0.3$ deg.
They also claimed that N1425 is the sdOB + dM eclipsing binary. 
\cite{2010Ap&SS.329..113Q} and \cite{2011ASPC..451..155Z} found the first hint of a cyclic period change in this system.  
\cite{2012MNRAS.421.3238K} discovered that the orbital period of N1425 is rapidly increasing at a rate of about $12 \cdot 10^{-12}$ days per orbit.
On the other hand, \cite{2012A&A...540A...8B} reported that there may be a giant planet with a mass of roughly 12~\mj\ in N1425. 

Moreover, \cite{2013ApJ...766...11A} revisited the O-C diagram of N1425 and explained the variations in \oc\ curve by the presence of two circumbinary bodies with masses of 8.1 \mj\ and 2.9 \mj. 
\cite{2013MNRAS.431.2150W} presented a dynamical analysis of the orbital stability of the model suggested by \cite{2013ApJ...766...11A}. They found that the two-planet model in N1425 is unstable on timescales of less than a thousand years. 
Later, \cite{2014MNRAS.438..307H} concluded that the insufficient coverage of timing data prevents the reliable constrain. 

Next, \cite{2017AJ....153..137N} published  times of minimum light and extended the time interval up to 2016. They ruled out the two-planet model and reported a cyclic change that was explained as the presence of a brown dwarf. However, their data still do not cover a full orbital cycle.

Recently, \cite{2019RAA....19..134Z}, in their comprehensive period study based on numerous new mid-eclipse times, claimed that cyclic change detected in N1425 could be explained by the LITE caused by the presence of a third body. The minimal mass was determined as 14.15~\mj\ close to a giant planet or a brown dwarf. There is also a long research history of this unique object summarized in the last mentioned paper.

Table~\ref{m1425} contains new mid-eclipse times obtained mostly in \ond. Several earlier eclipse measurements were obtained at Masaryk University Observatory  (MUO) during the summer of 2009. The 0.6-m reflecting telescope and the CCD camera SBIG ST-8 were used.
One additional eclipse light curve of N1425 and precise mid-eclipse time
was obtained by PZ at San Pedro M\'artir Observatory, UNAM, Baja California, Mexico, with the 0.84-m telescope in August 2009 (JD 24~55050).
The following linear light elements were adopted for calculation of epochs, other columns are self-explanatory:

\begin{center}
Pri.Min. = BJD 24 54274.20917 + 0\fd110374106 $\cdot\ E$.
\end{center}

\begin{figure}[t!]
\centering
\includegraphics[width=\columnwidth]{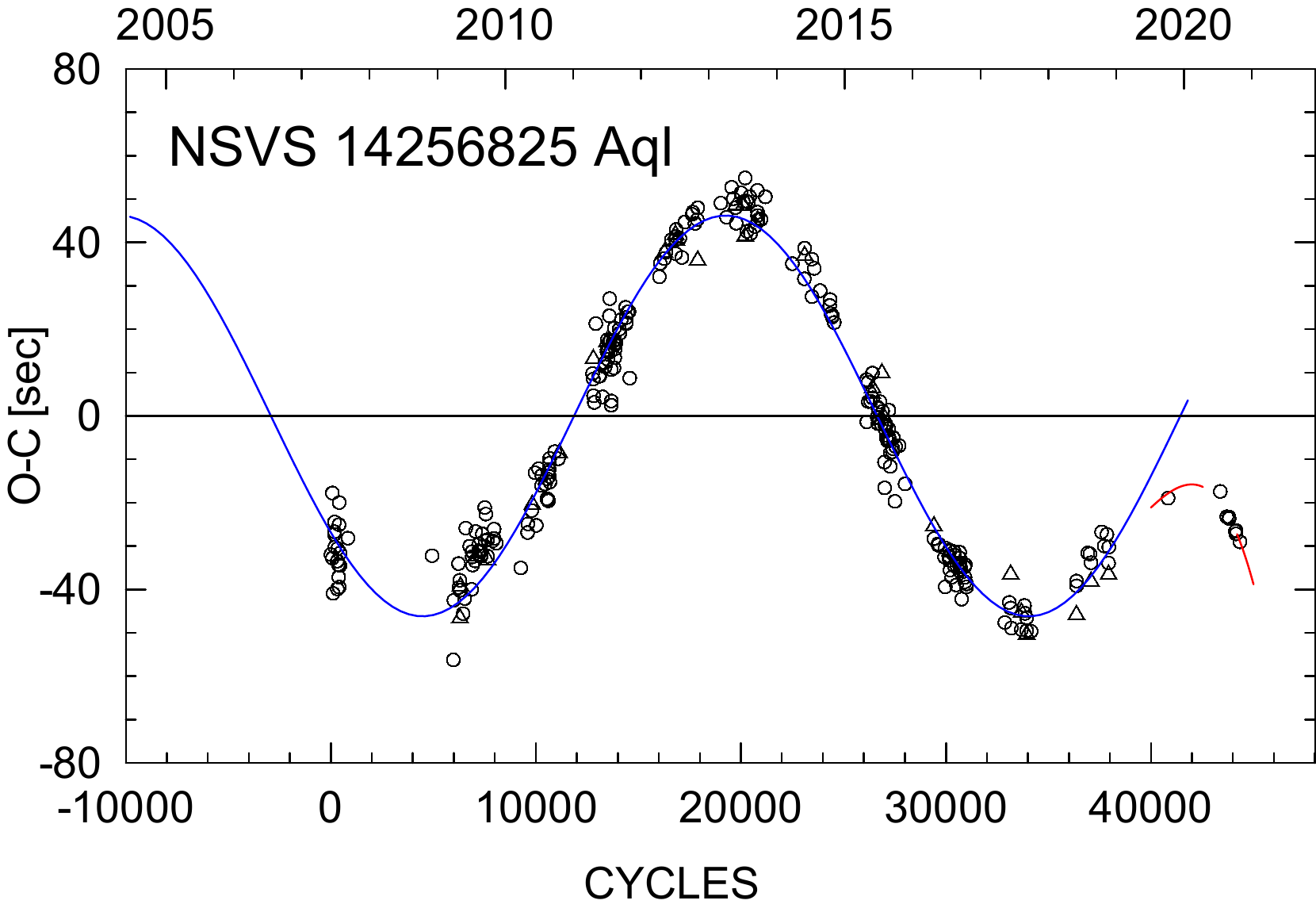}
\caption{Actual O-C diagram for the eclipse times of N1425 (the last minimum obtained in November~2020). The individual primary minima are denoted by circles, the secondary by triangles. The blue sinusoidal curve with the short 
        period of about 14~years clearly fits all data up to the epoch 40~000 (September~2019), but it does not follow the last mid-eclipse times measured after this date (red dashed curve).
        }
\label{n1425}
\vspace{5mm}
\includegraphics[width=\columnwidth]{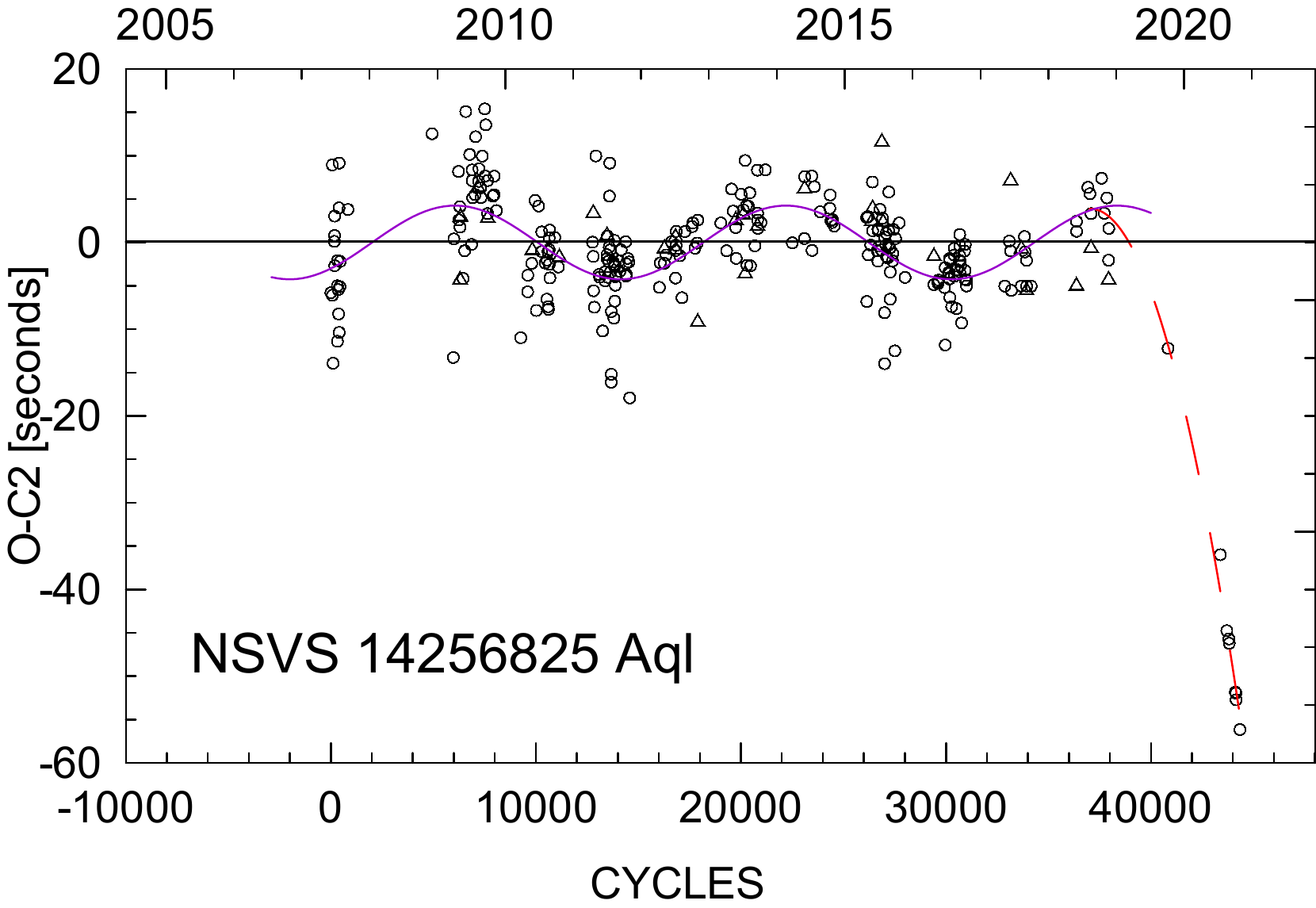}
\caption[]{ O-C2 diagram for N1425 after subtraction of a sinusoidal term
    of the possible third body. The violet curve represents
    additional cyclic variations with a period of about 5 years and an amplitude of 4 sec. The present rapid period decrease is denoted by red dashed curve.
                }
\label{n1425-3}
\end{figure}

\begin{table}[t]
\caption{New minima timings of N1425.} 
\label{m1425}
\begin{tabular}{llrc}
\hline\hline\noalign{\smallskip}
BJD$_{\rm TDB}$ -- &  Error & Epoch  & Weight   \\
24 00000          &  [day] &        &          \\
\noalign{\smallskip}\hline
\noalign{\smallskip}
  55021.44152*  &  0.0001  &  6770.0   & 5   \\
  55034.46565*  &  0.0001  &  6888.0   & 5   \\
  55034.57601*  &  0.0001  &  6889.0   & 5   \\
  55050.911367**&  0.00001 &  7037.0   & 10  \\
  55053.45005*  &  0.0001  &  7060.0   & 5   \\
  55358.468971  &  0.0001  &  9823.5   & 0   \\
  55373.424740  &  0.00001 &  9959.0   & 10  \\
  55392.409097  &  0.00001 &  10131.0  & 10  \\
  55784.347812  &  0.00001 &  13682.0  & 10  \\
  55796.37871*  &  0.00001 &  13791.0  & 5   \\
  55796.48924*  &  0.00001 &  13792.0  & 5   \\
  56107.413167  &  0.00001 &  16609.0  & 10  \\
  56133.351109  &  0.00001 &  16844.0  & 10  \\
  56179.377131  &  0.00001 &  17261.0  & 10  \\
  56482.574878* &  0.00001 &  20008.0  & 10  \\
  56494.440068  &  0.0001  &  20115.5  & 0   \\
  59494.495258  &  0.00001 &  20116.0  & 10  \\
  56529.373489  &  0.00001 &  20432.0  & 10  \\
  56613.257809  &  0.00001 &  21192.0  & 10  \\
  56824.458511  &  0.0001  &  23105.5  & 0   \\
  56824.513711  &  0.00001 &  23106.0  & 10  \\
  56876.499861  &  0.00001 &  23577.0  & 10  \\
  57219.542209  &  0.00001 &  26685.0  & 10  \\
  57230.358889  &  0.00001 &  26783.0  & 10  \\
  57240.347829  &  0.0001  &  26873.5  &  0  \\
  57297.355838  &  0.00001 &  27390.0  & 10  \\
  57616.336669  &  0.00001 &  30280.0  & 10  \\
  57645.365088  &  0.00001 &  30543.0  & 10  \\
  57696.247496  &  0.00001 &  31004.0  & 10  \\ 
  58044.256934  &  0.00001 &  34157.0  & 10  \\
  58349.331172  &  0.00002 &  36921.0  & 10  \\
  58423.281879  &  0.00001 &  37591.0  & 10  \\
  58437.188979  &  0.00001 &  37717.0  & 10  \\ 
  58781.335569  &  0.00001 &  40835.0  & 10  \\
  59062.458435  &  0.00001 &  43382.0  & 10  \\
  59097.336585  &  0.00001 &  43698.0  & 10  \\
  59108.373995  &  0.00001 &  43798.0  & 10  \\
  59110.360725  &  0.00001 &  43816.0  & 10  \\ 
  59143.252174  &  0.0001  &  44114.0  & 5   \\
  59147.225634  &  0.00001 &  44150.0  & 10  \\
  59168.196694  & 0.00001  &  44340.0  & 10 \\
\noalign{\smallskip}\hline
\end{tabular}
\tablefoot{* MUO, Brno, Czech Republic, 
** San Pedro M\'artir Observatory, UNAM, Mexico. 
            }
\end{table}

\noindent
A total of 310 mid-eclipse times were included to our analysis. As in a previous analysis, the secondary minima were not included due to their lower accuracy.

The corresponding O-C diagram is plotted in Fig.~\ref{n1425}, where the cyclical change with a period of about 14 years is clearly visible. 
The best fit is plotted as a continuous blue curve.
As one can see, the mid-eclipse times after the epoch 40~000 (September 2019) do not follow the predicted sinusoidal trend and a rapid period decrease is significant.
 The O-C2 diagram after subtraction of previous sinusoidal term is plotted in Fig.~\ref{n1425-3}. Additional cyclic variations with a period of 5 years and a small amplitude of about 4 sec are remarkable in these residuals.
Thus, a multiple companion system or more complicated process caused by as yet unknown effects can likely explain the period changes of this PCEB binary. In any case, a single third body orbiting the eclipsing pair is not sufficient to describe the current shape of the O-C diagram. 

\begin{table}
\caption{Physical properties of S1435 and N782 and parameters  of their possible third bodies.}
\begin{center}
\begin{tabular}{cccc}
\hline\hline\noalign{\smallskip}
Parameter & Unit     &  S1435      &  N782       \\
\hline\noalign{\smallskip}
$M_1$       & \ms    & 0.50(2)     &  0.442(12)  \\
$M_2$       & \ms    & 0.21(3)     &  0.124(5)   \\
$R_1$       & \rs    & 0.0145(2)   &  0.172(2)   \\
$R_2$       & \rs    & 0.23(2)     &  0.157(2)   \\
\hline\noalign{\smallskip}
Source &  & \cite{2009MNRAS.394..978P} & \cite{2017ApJ...839...39L}  \\
\hline\noalign{\smallskip}      
$f(m_3)$        & \ms & $7.6\cdot 10^{-6}$ & $6.0\cdot 10^{-9}$   \\
$M_{3,\rm min}$ & \ms &  0.016       &  0.0013        \\
$M_{3,\rm min}$ & \mj &  16.7        &  1.36          \\
$K$             & km/s &  0.25      &  0.025         \\
$A_{\rm dyn}$   & day  & $3.4\cdot 10^{-8}$ & $2.8 \cdot 10^{-6}$ \\
\hline\noalign{\smallskip}
\end{tabular}
\end{center}
\label{t3}
\end{table}

\section{Discussion}

The discovery of the LITE also allows us to determine
the stellar multiplicity in dwarf binary stars.
The derived third-body parameters lead to the following equation for the mass function
\citep{1990BAICz..41..231M}:

\medskip
\noindent
$$ f(M) = \frac{M_3^3 \sin^3 i_3}{(M_1+M_2+M_3)^2} 
        = \frac{1}{P^2_3} \, \left[ \frac {173.15 \, A} 
         {\sqrt{1 - e_3^2 \cos^2 \omega_3}} \right] ^3, $$

\smallskip
\noindent 
where $M_i$ are the masses of components. The systemic radial velocity of the eclipsing pair has an amplitude (in km/s) of
$$ K = \frac{A}{P_3} \frac{5156}{\sqrt{\left(1-e^2_3\right)\,\left(1-e^2_3 \cos^2 \omega_3\right)}}. $$

\smallskip
\noindent
Assuming a coplanar orbit of the third component ($i_3 \sim 90^{\circ}$), we can obtain a lower limit for its mass $M_{3, \rm min}$. 
These quantities for the third body of individual systems are collected in Table~\ref{t3}. 
The amplitude of the dynamical contribution of the third body, 
$A_{\rm dyn}$, is given by \citep{2016MNRAS.455.4136B}

$$ A_{\rm dyn} = \frac{1}{2\pi} \frac{M_3}{M_1+M_2+M_3} \frac{P_s^2}{P_3} 
             \, \left(1-e^2_3\right)^{-3/2} $$

\smallskip
\noindent
and is also listed in Table~\ref{t3}. The value of $A_{\rm dyn}$ is of the order of tenths of a second and is less than an individual mid-eclipse time precision.

Another possible mechanism for cyclical period variation frequently discussed in literature is a magnetic activity cycle for systems with a late-type secondary component \citep{1992ApJ...385..621A}. 
Recently, \cite{2018A&A...615A..81N} showed that the Applegate mechanism is energetically feasible in five PCEB systems. However, for N1425 they note that there are no solutions that could explain the eclipsing time variations entirely, but magnetic activity could at least induce relevant scatter in the observed variations. 

For N1425 and N782, we also used the publicly available Eclipsing time variation 
calculator\footnote{Applegate calculator: \\
\url{http://theory-starformation-group.cl/applegate/}.}
based on the two-zone model by \cite{2016A&A...587A..34V}.  For the updated parameters in Table~\ref{t2}, we find that the required energy to drive the Applegate mechanism is approximately $10^2 - 10^5$ times the available energy in the magnetically active secondary (solution for the finite-shell two-zone model). 
The newly derived LITE periods for selected objects 
($\sim$ 10~years) are too short for the magnetic cycle, and 
this mechanism cannot contribute significantly to the 
observed period changes in these systems. 
In case of S1435 with a minute amplitude of variations, the above-mentioned calculator only gives a physical solution for the finite-shell constant density model ($\Delta E/E_{sec} \approx 10^7$).

The long-term eclipse timings of white dwarf binaries with respect to a magnetic mechanism was presented by \cite{2016MNRAS.460.3873B}. 
They found that all binaries with baselines exceeding 10~years, with secondaries of spectral type K2 -- M5.5, show variations in the eclipse arrival times that in most cases amount to several minutes. They conclude that a still relatively short observational baseline for many of the binaries cannot yet provide obvious conclusions about the cause of orbital period variations in white dwarf binaries. 

The stability of circumbinary companions in post-common envelope eclipsing binaries was discussed by \cite{2018A&A...611A..48P}. They conclude that period variation cannot be modeled simply on the basis of a circumbinary object, thus a more complex processes may be taking place in some systems.

In case S1435, we note the unseen third body orbiting the eclipsing pair with a mass of about 17 \mj\  which is in transition from planet
to brown dwarf, but is well-below the stellar-mass limit of 0.075 \ms. 
On the other hand, such an orbiting body could be confirmed
spectroscopically using modern instruments connected 
to medium-sized telescopes. The derived amplitude of the systemic radial velocity of S1435 is about 250 m/s (see Table~\ref{t3}). Finally, in N782 we expect a giant planet of Jupiter mass. 

 A similar abrupt change in the eclipse timings as observed in N1425 also occurred in the well-known PCEB system HS~0705+6700 \citep[][]{2018A&A...611A..48P}. This system underwent a sudden extension of the period in February 2015 (see their Figs. 3 and 4). 
Thus, the previous single third-body hypothesis can no longer be valid.
For such systems, we can introduce a two-satellite model, in which one body is orbiting the eclipsing pair on a nearly circular orbit, whilst the second companion has a highly eccentric and long-period orbit and has just passed through the periastron. After this passage, the system should relax to previous state with nearly the same periodicity.
On the other hand, two circumbinary brown dwarfs with orbital periods of about 8 and 13 years were recently
proposed for HS~0705+6700
\citep{2020MNRAS.499.3071S}, with a relatively good fit of eclipse times, but in dynamically unstable configuration.

\section{Conclusions}

Our careful analysis of O-C diagrams has provided identification or confirmation of the two probable triple systems between known dwarf binaries, namely S1435 and N782. In both systems, the whole third-body orbital period is now measured by the reliable mid-eclipse times. The cyclic variations of the orbital period are explained by the LITE caused by a third body as the more probable scenario,
most likely a brown dwarf or a giant planet with a mass of about 17~\mj\ in S1435 or 1.4 \mj\ in N782.

In the case of N1425, the previous LITE solution supported by many investigators is not confirmed by current timings.
We cannot approve a single additional body with the orbital period of about 8.8~years, as was last announced by \cite{2019RAA....19..134Z}.
For this system, we propose an explanation using at least two additional bodies. A longer time span is required for an accurate multiple-satellite solution.

Future observations of these interesting objects could offer a more concrete  explanation for their period changes, which could be caused by a currently unknown or unexpected phenomenon connected with the internal structure of the components,  an evolutionary effect, or circumbinary bodies. The sample of well-known PCEB or sdB binaries needs to be increased, and observations of additional systems would be very useful. 

\medskip
\begin{acknowledgements}
Useful suggestions and recommendation by an anonymous referee helped to improve the clarity of the paper and are greatly appreciated.
M.W. was supported by the Czech Science Foundation grant GA19-01995S. The research of M.W. and P.Z. was also supported by the project Progress Q47 {\sc Physics} of the Charles University in Prague.
H.K. and K.H. were supported by the project RVO: 67985815.
The authors would also like to thank 
Lenka Kotkov\'a, \ond\ observatory, 
Jan Vra\v{s}til, Charles University in Prague, 
Ji\v{r}\'i Li\v{s}ka and Marek Chrastina, Masaryk University Brno, 
Ladislav \v{S}melcer, \valmez\ observatory, 
Reinhold Friedrich Auer, S-M-O \vev\ observatory, 
all from the Czech Republic,
for their important contribution to photometric observations.
This paper includes data collected by the TESS mission. Funding for the TESS mission is provided by the NASA Explorer Program.
The following internet-based resources were used in research for this paper: the SIMBAD database and the VizieR service operated at CDS, Strasbourg, France;
the NASA's Astrophysics Data System Bibliographic Services.
This work has made use of data from 
the European Space Agency (ESA) mission
{\it Gaia} (\url{https://www.cosmos.esa.int/gaia}), 
processed by the {\it Gaia}
Data Processing and Analysis Consortium (DPAC,
\url{https://www.cosmos.esa.int/web/gaia/dpac/consortium}). 
Funding for the DPAC has been provided by national institutions, 
in particular the institutions participating in the {\it Gaia} Multilateral Agreement.
This research is part of an ongoing collaboration between professional 
astronomers and the Czech Astronomical Society, Variable Star 
and Exoplanet Section.

\end{acknowledgements}

\bibliographystyle{aa.bst}
\bibliography{s1435.bib}

\end{document}